%% file: main.tex
\documentclass{Interspeech2024}

\interspeechcameraready 

\title{FlowAVSE: Efficient Audio-Visual Speech Enhancement \\ with Conditional Flow Matching}

\name{Chaeyoung}{Jung}
\name{Suyeon}{Lee}
\name{Ji-Hoon}{Kim}
\name{Joon Son}{Chung}

\address{
  Korea Advanced Institute of Science and Technology, South Korea
  }
\email{\{codud9914, syl4356, jh.kim, joonson\}@kaist.ac.kr}


\begin{document}

\maketitle

\begin{abstract}
This work proposes an efficient method to enhance the quality of corrupted speech signals by leveraging both acoustic and visual cues.
While existing diffusion-based approaches have demonstrated remarkable quality, their applicability is limited by slow inference speeds and computational complexity.
To address this issue, we present FlowAVSE which enhances the inference speed and reduces the number of learnable parameters without degrading the output quality.
In particular, we employ a conditional flow matching algorithm that enables the generation of high-quality speech in a single sampling step.
Moreover, we increase efficiency by optimizing the underlying U-net architecture of diffusion-based systems. 
Our experiments demonstrate that FlowAVSE achieves 22 times faster inference speed and reduces the model size by half while maintaining the output quality.
The demo page is available at: \url{https://cyongong.github.io/FlowAVSE.github.io/}

\end{abstract}

\input{Contents/Intro}

\input{Contents/Method}

\input{Contents/Experiments}

\input{Contents/Conclusion}

\input{Contents/ack}

\bibliographystyle{IEEEtran}
\bibliography{main}

\end{document}

%% file: Contents/Intro.tex
\section{Introduction}
\label{sec:intro}

Despite recent advancements in audio-based speech enhancement systems, significant challenges remain in extracting clean speech from noisy environments, often resulting in residual background noise in the enhanced audio.
Interestingly, humans tend to understand spoken language more effectively in face-to-face interactions than in telephonic conversations~\cite{looking,c29}.
This observation highlights the importance of incorporating visual cues alongside auditory data. Similarly, speech enhancement systems with visual information achieve better results than the audio-only approaches~\cite{looking,c29, rahimi2022reading,richter2023audio,owens2018audio,lai2023audio}.

As a result, various Audio-Visual Speech Enhancement (AVSE) systems have been developed. These systems aim to isolate clear speech by leveraging both sound and visual information, thereby opening up new applications such as denoising for video conferencing to meet the increasing demand for online meetings.
Existing works in AVSE can be broadly classified into two categories: predictive and generative approaches. 
Initially, the focus was on the predictive approach that directly predicts clean speech or the spectrogram mask by reducing spectrogram differences between clean and predicted speech~\cite{looking,c29,visualvoice}.
While these methods demonstrate the advantages of incorporating visual and acoustic cues, they still encounter issues with over-denoising, which can result in unnatural speech output~\cite{storm}.

\begin{figure}[!t]
    \centering
    \includegraphics[width=0.9\linewidth]{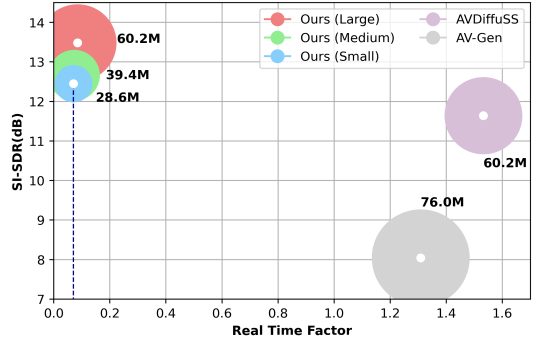}
    \vspace{-3mm}
    \caption{Analysis of inference speed, parameter size, and SI-SDR scores on VoxCeleb2 test set. The real-time factor measures the time needed for 1 second of audio generation. Our small size model showcases an inference speed approximately 22 times faster and 2 times lighter than the previous model while achieving superior performance.}
    \label{fig1}
    \vspace{-1mm}
\end{figure}

On the other hand, generative methods have shown promise in producing high-quality speech. 
The work of~\cite{sadeghi2020audio} shows promising results based on variational autoencoder~\cite{kingma2014}, and the following work~\cite{yang2022audio} exploits speech codecs~\cite{van2017neural} to improve the quality.
However, these approaches still face difficulties in generating diverse and natural-sounding speech. 
Diffusion models~\cite{ho2020denoising, song2019generative}, known for their ability to generate various and high-quality samples, have demonstrated strong results across different domains, including image~\cite{nichol2022glide, latentdif, saharia2022photorealistic}, video~\cite{ho2022video, ho2022imagen}, and speech~\cite{popov2021grad, sgmse+} generation.
However, their applicability is limited by slow inference speed due to the multi-step inference process.
Such a problem is a significant drawback for real-time applications and there have been several attempts to resolve the problem~\cite{song2020denoising,salimans2022progressive} by using diffusion models.

To address a slow inference speed issue of diffusion-based approach, we present \textbf{FlowAVSE}, a novel \textbf{A}udio-\textbf{V}isual \textbf{S}peech \textbf{E}nhancement model based on conditional \textbf{Flow} matching. FlowAVSE delivers outstanding performance with fast inference speeds and low memory usage. Employing a conditional flow matching model~\cite{lipman2022flow,tong2023improving} circumvents the necessity for initialization from a standard normal distribution, enabling the denoising of samples in a single inference step. This represents a significant efficiency over the 30-step sampling procedure required by diffusion-based models~\cite{richter2023audio,storm,lee2023seeing}. 
Furthermore, diffusion-based speech models~\cite{storm,sgmse+, lee2023seeing} typically depend on a U-net architecture known as the Noise Conditional Score Network (NCSN) originally for image synthesis~\cite{song2019generative}, but its large size poses a challenge during training and inference phases.
FlowAVSE streamlines this framework by reducing repeated components from the architecture.  
Experimental results demonstrate that our model significantly boosts computational efficiency while generating natural, high-quality output. 

In summary, we present the first audio-visual speech enhancement system based on a conditional flow matching model, capable of fast inference. We also propose a refined NCSN architecture that decreases model complexity for a minimal compromise in performance. As shown in Fig.~\ref{fig1}, FlowAVSE achieves 22 times faster inference speed and half the number of parameters compared to the previous diffusion-based model, while achieving superior quality in terms of SI-SDR.

%% file: Contents/Method.tex
\section{Method}
\label{sec:method}

\subsection{Proposed architecture}
\label{ssec:method_proposed}
\begin{figure}[!t]
  \centering
  \includegraphics[width=\linewidth]{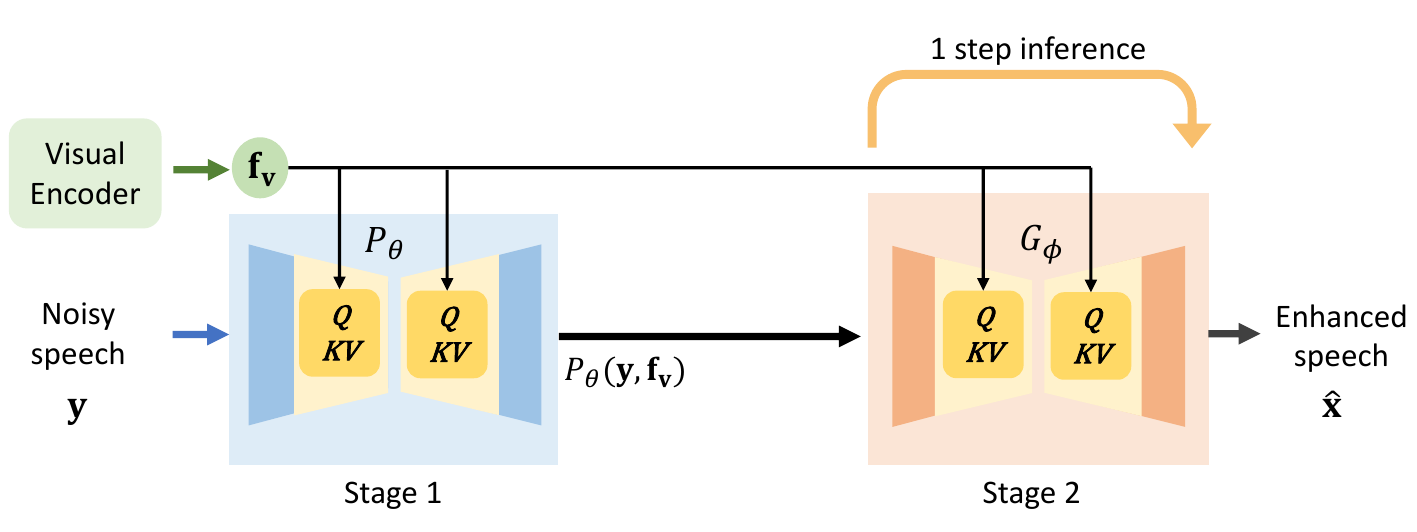}
  \caption{Model architecture of FlowAVSE. Face-cropped images of the speaker are fed to the visual encoder to acquire visual embedding $\mathbf{f_v}$. Through the $P_\theta$ and $G_\phi$, visual embedding $\mathbf{f_v}$ is fused with auditory information from the noisy speech $\mathbf{y}$ to obtain an enhanced speech $\hat{\mathbf{x}}$. Both stages consist of U-net architecture and are trained simultaneously by $\mathcal{L}_{total}$.
  }
  \label{fig2}
\end{figure}

In Fig.~\ref{fig2}, our framework consists of two primary phases. During the first stage $P_\theta$, the model predicts the speech from the noisy speech $\mathbf{y}$ by using visual semantics $\mathbf{f_v}$ obtained from the visual encoder. Taking insight from a study in active speaker detection~\cite{talknet, jung2023talknce}, a visual encoder with the ability to retain temporal dynamics can be leveraged, which is jointly optimized with $P_\theta$ and $G_\phi$. The resulting output of the first stage represented as $P_\theta(\mathbf{y, f_v})$, is subsequently passed to the second stage $G_\phi$, where a conditional flow matching model is employed. Through the secondary stage, the output of the first stage undergoes further refinement. Both stages include cross-attention modules similar to AVDiffuSS~\cite{lee2023seeing} to temporally align the visual embedding $\mathbf{f_v}$ with the auditory information in our light U-net architectures, detailed in Section~\ref{ssec:method_unet}.

\subsection{Flow matching}
\label{ssec:method_fm}

\input{Tables/SEtable}

Conditional flow matching~\cite{lipman2022flow, tong2023improving} is utilized to train $G_\phi$ in our model to refine the output of the first stage
and accelerate the inference speed at the same time. 
Flow matching is a method for training Continuous Normalizing Flows (CNFs)~\cite{chen2018neural}. CNFs are continuous-time versions of normalizing flows~\cite{rezende2015variational}, which generate samples by invertible mapping between two distributions.
To sample data $\mathbf{x} \in \mathbb{R}^d$ from the ground-truth data distribution $q(\mathbf{x})$, we approximate $q(\mathbf{x})$ by exploiting a time-dependent probability density path $p_t: [0,1] \times \mathbb{R}^d \rightarrow \mathbb{R}_{>0}$, where $t \in [0,1]$. 
Starting from the prior distribution $p_0(\mathbf{x}) = \mathcal{N}(\mathbf{x}; 0, \mathbf{I})$ at $t=0$, $p_t$ is designed to  approximate the data distribution $q(\mathbf{x})$ as $t \rightarrow 1$.
A time-dependent flow $\psi_t : [0,1] \times \mathbb{R}^d \rightarrow \mathbb{R}^d$, which produces $p_t$, can be generated by a time-dependent vector field $\mathbf{v}_t: [0,1] \times \mathbb{R}^d \rightarrow \mathbb{R}^d$, defined by the following Ordinary Differential Equation (ODE):
\begin{equation}
    \frac{d}{d t} \psi_t(\mathbf{x})=\mathbf{v}_t\left(\psi_t(\mathbf{x})\right),
\end{equation}
where the initial condition is given as $\psi_0(\mathbf{x})=\mathbf{x}$.

Let $\mathbf{u}_t$ a target vector field that produces a probability path $p_t$ from $p_0$ to $p_1 \approx q$. It is impractical to directly compute the vector field $\mathbf{u}_t$ and the target probability path $p_t$ as they are intractable, so \cite{tong2023improving} resolved the problem by conditioning on $\mathbf{z}$.
By using the conditional vector fields $\mathbf{u}_t(\mathbf{x}|\mathbf{z})$ and conditional probability paths $p_t(\mathbf{x}|\mathbf{z})$, conditional flow matching (CFM) loss is suggested for the regression of marginal vector field $\mathbf{u}_t\left(\mathbf{x}\mid\mathbf{z}\right)$ as follows:
\begin{equation}
   \mathcal{L}_{\mathrm{CFM}}(\phi)=\mathbb{E}_{\scaleto{t, q(\mathbf{z}), p_t(\mathbf{x}|\mathbf{z})}{7pt}} {||\mathbf{v}_\phi(\mathbf{x, t}) - \mathbf{u}_t\left(\mathbf{x}\mid\mathbf{z}\right)||}^2,
\label{condi}
\end{equation}
where $t$ is sampled from a uniform distribution $t \sim \mathbf{U}[0,1]$, $\mathbf{z}$ is sampled from the distribution $q(\mathbf{z})$, $\mathbf{x}$ is sampled from the conditional distribution $p_t(\mathbf{x}|\mathbf{z})$, and $\mathbf{v}_\phi(\mathbf{x}, t)$ represents a neural network with parameters $\phi$.

Given the time-dependent Gaussian conditional path $
p_t(\mathbf{x}|\mathbf{z}) = \mathcal{N}(\mathbf{x} |\mathbf{\mu_t(\mathbf{z})},\sigma_t(\mathbf{\mathbf{z}})^{2}\mathbf{I})$,
the flow $\psi_t$ could be constructed simply as follows:
\begin{equation}
    \psi_t(\mathbf{x}) = \sigma_t(\mathbf{z})\mathbf{x}+\mu_t(\mathbf{z}).
\label{flowmap}
\end{equation}

Following \cite{lipman2022flow, tong2023improving}, the unique vector field that generates the flow $\psi_t$ is as follows:
\begin{equation}
    \mathbf{u}_t\left(\mathbf{x} \mid \mathbf{z}\right)=\frac{\sigma_t^{\prime}\left(\mathbf{z}\right)}{\sigma_t\left(\mathbf{z}\right)}\left(\mathbf{x}-\mu_t\left(\mathbf{z}\right)\right)+\mu_t^{\prime}\left(\mathbf{z}\right),
\end{equation}
where $\sigma_t^{\prime}$ and $\mu_t^{\prime}$ are the time derivatives of $\sigma_t$ and $\mu_t$.

\newpara{Simplified CFM.}
Conditional flow matching objectives can be chosen with any conditional probability path and vector fields. In the simplified version of CFM~\cite{tong2023improving}, the inference stage starts with non-standard normal distribution for better performances.
Simplified CFM defines the condition as $\mathbf{z}:= (\mathbf{x}_0, \mathbf{x}_1)$, which are tuple points from the joint distribution $(q_0, q_1)$. Data samples $\mathbf{x}_0$ and $\mathbf{x}_1$ are sampled from $q_0$ and $q_1$, respectively. 
Conditional probability density path $p_t(\mathbf{x}|\mathbf{z})$ is set as a Gaussian distribution with a mean that is a linear interpolation between $\mathbf{x}_0$ and $\mathbf{x}_1$, with a fixed standard deviation value of $\sigma$. 
Therefore, the conditional probability path and vector field used for the simplified CFM can be described as follows:

\begin{equation}
    q(\mathbf{z}) := q(\mathbf{x_0})q(\mathbf{x_1}),
\end{equation}
\begin{equation}
    p_t(\mathbf{x}|\mathbf{z}) = \mathcal{N}(\mathbf{x} | t \mathbf{x_1}+(1-t)\mathbf{x_0},\sigma^{2}),    
\label{eq:pt}
\end{equation}
\begin{equation}
   \mathbf{u}_t(\mathbf{x}|\mathbf{z}) = \mathbf{x_1} -\mathbf{x_0}. 
\end{equation}

At the starting point $t=0$, $p_0(\mathbf{x}|\mathbf{z})$ is $ \mathcal{N}(\mathbf{x} | \mathbf{x_0},\sigma^{2})$ which corresponds to the Gaussian distribution centered at $\mathbf{x_0}$ and we use $P_\theta(\mathbf{y,f_v})$ as a $\mathbf{x_0}$. As $t \rightarrow 1$, the mean of $p_t$ approaches $\mathbf{x}_1$, which corresponds to the desired clean speech in our task.

Consequently, we set a simplified CFM loss as follows:
\begin{equation}
    \resizebox{0.91\hsize}{!}{%
  $\mathcal{L}_{\mathrm{CFM_{sim}}}(\phi)=\mathbb{E}_{\scaleto{t, q(\mathbf{z}), p_t(\mathbf{x}|\mathbf{z})}{7pt}} ||\mathbf{v}_\phi(\psi_t(\mathbf{x}_0), t)\! - (\mathbf{x}_1\! - \mathbf{x}_0)||^2,$
  }
\label{eq_CFM}
\end{equation}
where the second stage $G_\phi$ in our model is trained to play the role of the vector field $\mathbf{v}_t$.

\newpara{Training objective.}
The first stage, $P_\theta$, and the second stage, $G_\phi$, are jointly trained using a multi-task learning strategy, as outlined in~\cite{lee2023seeing}.
In the training process, the predictor $P_\theta$ learns to separate the target speech from $\mathbf{y}$ using the visual semantics $\mathbf{f_v}$. The loss function $\mathcal{L}_{p}$ for $P_\theta$ is computed using the MSE loss function between the initial prediction $P_\theta(\mathbf{y, f_v})$ and the ground-truth $\mathbf{x_1}$.
In Eq.~(\ref{eq_CFM}), we set $\mathbf{x_0}$ as $P_\theta(\mathbf{y},\mathbf{f_v})$ and $\mathbf{x_1}$ as a ground truth.  
To ensure balanced training, weight values $\lambda_1$ and $\lambda_2$ are assigned to $\mathcal{L}_p$ and $\mathcal{L}_\mathrm{CFM_{sim}}$, respectively. The total loss $\mathcal{L}_{total}$ is then computed as follows:
\begin{equation}
    \mathcal{L}_{p}(\theta) = \mathbb{E} \left[ || \mathbf{x_1} - P_\theta(\mathbf{y, f_v})||^2_2 \right], \label{eq:predloss}
\end{equation}
\begin{equation}
    \mathcal{L}_{total} = \lambda_1 * \mathcal{L}_{p}(\theta) + \lambda_2 * \mathcal{L}_{\mathrm{CFM_{sim}}}(\phi). \label{eq:totalloss}
\end{equation}

\begin{figure}[!t]
  \centering
  \includegraphics[width=0.95\linewidth]{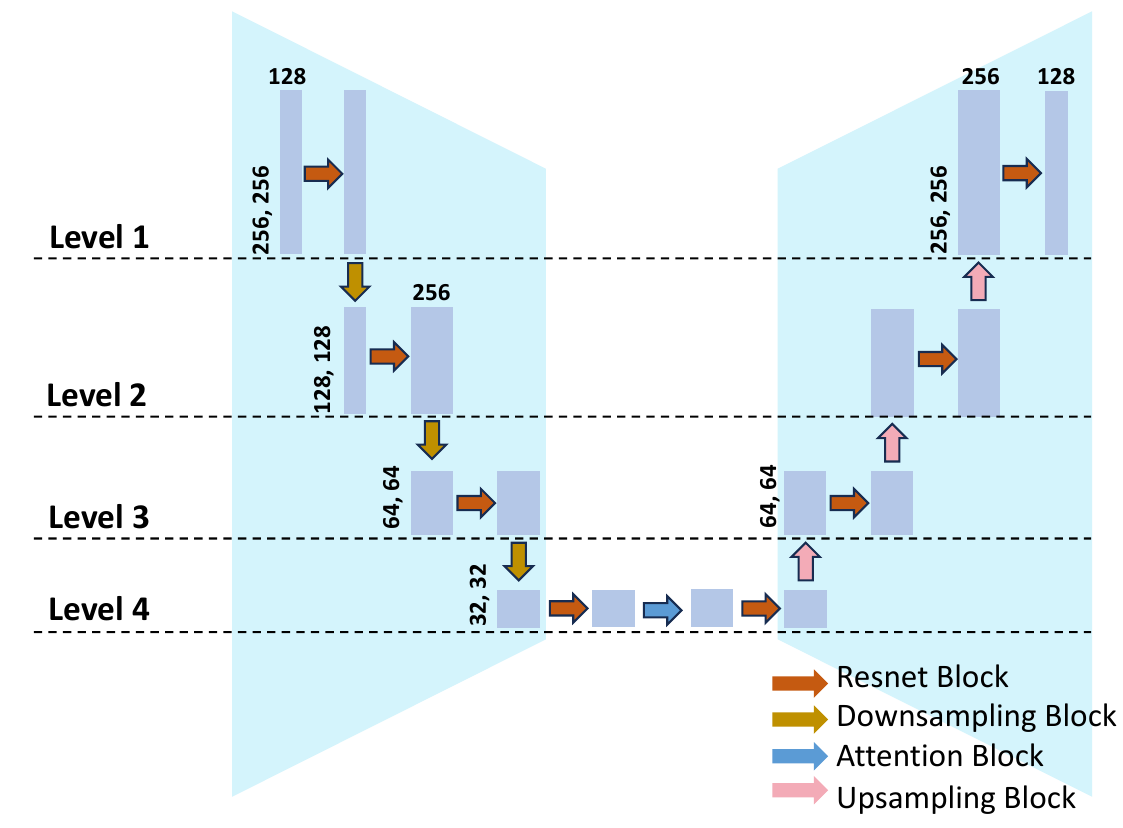}
  \caption{A simplified illustration of the U-net architecture in our model. We remove duplicate convolution modules for enhanced efficiency in our small and medium size models.}
  \label{fig3}
\end{figure}

\subsection{Light U-net}
\label{ssec:method_unet}
To further boost efficiency, we optimize U-net architecture which has gained widespread adoption in diverse research areas~\cite{ronneberger2015u,lee2020mu}. 
As illustrated in Fig.~\ref{fig3}, a typical U-net architecture consists of four downsampling layers and four upsampling layers. During the downsampling stage, the input dimension is halved while the channel is doubled, which can facilitate the capture of contextual information. 
Conversely, the module accurately restores the dense maps by leveraging coarse maps from the downsampling stage and skip connections in U-net.

In this study, we adopt NCSN for both stages. 
NCSN is structured based on the U-net architecture and has demonstrated its effectiveness in diffusion models~\cite{song2019generative}.
Our goal is to identify the essential components of NCSN++M~\cite{storm}, which is a modified version of existing NCSN, and streamline its less important parts.
As the convolution modules are repeated in every block of NCSN++M, we hypothesize that simplifying those modules would enhance efficiency with minimal performance degradation. 
Therefore, we propose lighter versions of the model and justify our hypothesis through experiments.

%% file: Tables/SEtable.tex
\begin{table*}[!t]
\centering
\caption{Speech enhancement results on the VoxCeleb2 and LRS3 dataset. All models use audio-visual modalities. \textup{Steps} indicates the number of sampling steps. \textup{RTF} denotes real time factor indicating how much time is needed to generate one second of audio. For all metrics except for RTF, higher is better.} 
\resizebox{\linewidth}{!}{%
\begin{tabular}{lcccccrccr}
\toprule
                                 &    & & &\multicolumn{3}{c}{{VoxCeleb2}} & \multicolumn{3}{c}{{LRS3}} \\
                                  \cmidrule(lr){5-7} \cmidrule(lr){8-10} 
\multicolumn{1}{l}{Method}                       & Params (M) & Steps &RTF $\downarrow$       & PESQ $\uparrow$     & ESTOI $\uparrow$      & \multicolumn{1}{c}{SI-SDR$\uparrow$}      & PESQ $\uparrow$    & ESTOI $\uparrow$    & \multicolumn{1}{c}{SI-SDR$\uparrow$} \\
\midrule
\multicolumn{1}{l}{AV-Gen~\cite{richter2023audio}} &76.0 &30 &1.308   &1.690$\pm$0.001             &0.682$\pm$0.001 &8.056$\pm$0.001 & 1.892$\pm$0.016          & 0.782$\pm$0.004 & 9.618$\pm$0.151 \\ \midrule
\multicolumn{1}{l}{AVDiffuSS~\cite{lee2023seeing}}  &60.2 &30 &1.532   &\textbf{2.271$\pm$0.011}    &0.760$\pm$0.006 &11.508$\pm$0.132 & \textbf{2.271$\pm$0.040} & 0.831$\pm$0.011 & 12.587$\pm$0.007 \\ 
\multicolumn{1}{l}{AVDiffuSS~\cite{lee2023seeing}}  &60.2 &~~1 &0.083 &1.053$\pm$0.001   &0.076$\pm$0.001    &-17.272$\pm$0.058  & 1.035$\pm$0.001 & 0.107$\pm$0.001 & -16.746$\pm$0.008 \\ 
\midrule
\multicolumn{1}{l}{Ours (Small)}  &28.6 &~~1 &\textbf{0.070}   &2.053$\pm$0.023 &0.772$\pm$0.007            &12.341$\pm$0.109            & 1.962$\pm$0.051           & 0.830$\pm$0.004             & 13.265$\pm$0.131            \\
\multicolumn{1}{l}{Ours (Medium)}  &39.4 &~~1 &\underline{0.073}&2.011$\pm$0.045 &\underline{0.777$\pm$0.008}&\underline{12.457$\pm$0.208}& 1.975$\pm$0.069            & \underline{0.838$\pm$0.005} &\underline{13.561$\pm$0.117} \\ 
\multicolumn{1}{l}{Ours (Large)}  &60.2 &~~1 &0.085           &  \underline{2.096$\pm$0.027}    &    \textbf{0.796$\pm$0.006}      &    \textbf{13.370$\pm$0.111}    & \underline{2.077$\pm$0.066}& \textbf{0.850$\pm$0.006}    & \textbf{14.353$\pm$0.059}   \\
\bottomrule
\end{tabular}
}
\label{table1}
\end{table*}

%% file: Contents/Experiments.tex
\section{Experiments}
\label{sec:exp}

\begin{figure}[!t]
\vspace{-2mm}
  \centering
  \includegraphics[width=0.99\linewidth]{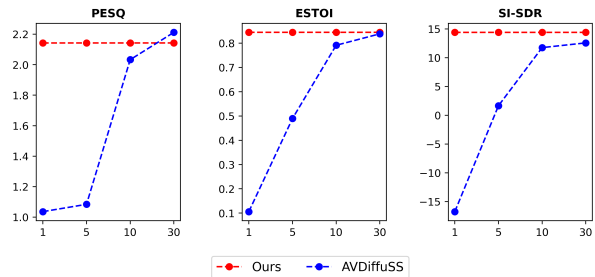}
  \vspace{-2mm}
  \caption{Comparison of AVDiffuSS and our model on the LRS3 test set across various sampling steps. Our model attains robust performances even with a single-step inference.}
  \vspace{-2mm}
  \label{fig_steps}
\end{figure}

In this section, we validate the effectiveness of FlowAVSE by using Perceptual Evaluation of Speech Quality (PESQ) \cite{pesq}, Extended Short-Time Objective Intelligibility (ESTOI)~\cite{estoi}, and Scale-Invariant Signal-to-Distortion Ratio (SI-SDR)~\cite{sisdr}.

\subsection{Datasets}
\label{ssec:dataset}
FlowAVSE is trained on VoxCeleb2~\cite{voxceleb2}, which is the established dataset for audio-visual speech enhancement. The training set consists of over 1 million speech segments and the test set is composed of 36,237 segments. 
We further evaluate FlowAVSE on the test set of the LRS3 dataset~\cite{afouras2018lrs3} to demonstrate the generalizability and robustness of our method. 
LRS3 contains 412 clips for the test sets. There is no speaker overlap between our training and test datasets.

We mix clean speech from VoxCeleb2 with the noise signal from AudioSet dataset~\cite{audioset} to construct noisy input speech.
AudioSet consists of various classes of audio samples, making it suitable for synthetic background noise.
The noise signal randomly chosen from AudioSet is mixed with the clean speech signal with an SNR value of 0 in both the train and test phases.

\subsection{Implementation details}
\label{ssec:details}
Visual encoder adopted from \cite{talknet} receives face-cropped images scaled to $112 \times 112$ and processes the input frames into visual embeddings through the 3-dimensional convolution layer followed by ResNet18~\cite{he2016deep}, temporal convolutional network~\cite{lea2017temporal}, and the final 1-dimensional convolution layer for compressing feature dimension. 
To update our model, we employ the Adam optimizer~\cite{KingBa15} with an exponential moving average of network parameters for stable training~\cite{song2020improved}, utilizing a decay rate of 0.999. The initial learning rate is set to $10^{-4}$.
For training, 4 RTX A5000 GPUs with a batch size of 16 are utilized. The training process extends for 15 epochs, spanning approximately two weeks. 
Training objective weight values $\lambda_1$ and $\lambda_2$ in Eq.~(\ref{eq:totalloss}) are empirically set to 0.5. For the CFM objective, the $\sigma$ value in Eq.~(\ref{eq:pt}) is fixed to $0.04$.
Pairs of clean and noisy speeches for performance evaluation are constructed using test sets of VoxCeleb2 and LRS3, following \cite{lee2023seeing}. 

To confirm the trade-off between efficiency and output quality, we conduct experiments as outlined in Table~\ref{table1} based on the three variations of our method: \textit{small}, \textit{medium}, and \textit{large}. 
The \textit{large} model adopts the original NCSN++M with cross-attention modules~\cite{lee2023seeing}, and \textit{medium} size model eliminates repeated convolution modules in NCSN++M.
Recognizing that the ResNet blocks in Level 4 of Fig.~\ref{fig3} are also repeated, we omit the inner two blocks in our \textit{small} model.

\begin{figure}[!t]
\vspace{-2mm}
  \centering
  \includegraphics[width=1.0\linewidth]{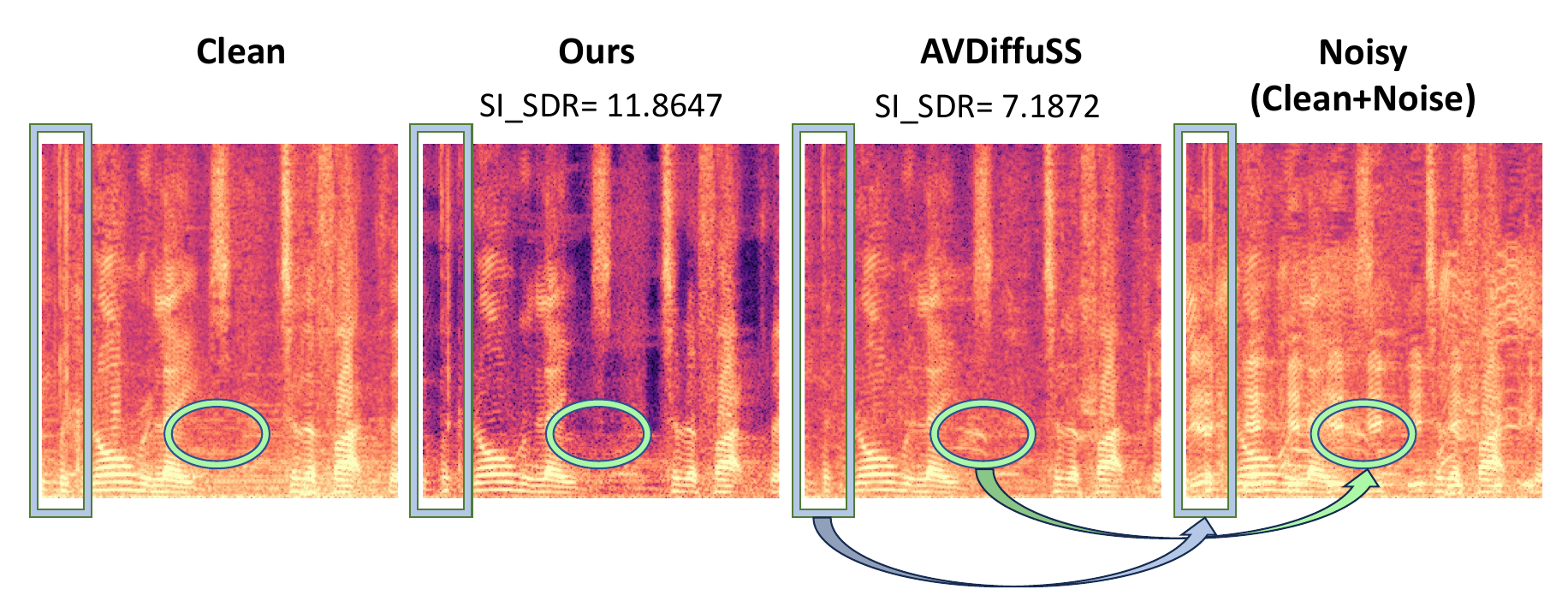}
  \vspace{-4mm}
  \caption{Comparison between the audio spectrograms of the diffusion-based model and ours. It shows that our model excels in removing not only the synthesized noise but also the background noise recorded along with the target speech.}
  \vspace{-2mm}
  \label{fig4}
\end{figure}

\subsection{Experimental results}
\label{ssec:results}
\newpara{Speech enhancement.}
We compare our method to the best performing audio-visual speech enhancement networks: AV-Gen~\cite{chern2023audio} and AVDiffuSS~\cite{lee2023seeing}.
Both are audio-visual diffusion models with different methods for incorporating visual information. AV-Gen leverages multi-modal embeddings from pre-trained AVHuBERT~\cite{shi2022learning} whereas AVDiffuSS incorporates visual modality through unique cross-attention modules.

As shown in Table~\ref{table1}, our model achieves state-of-the-art results across most of the evaluation metrics even with a single sampling step. 
In particular, our model exhibits significantly higher SI-SDR scores in spite of generative models, indicating the superior denoising capabilities of our method.
We further verify the effectiveness of FlowAVSE by visualizing the output spectrogram.
As illustrated in Fig.~\ref{fig4}, our model removes an inherent noise from the target speech as well as mixed synthetic noise.
This could bring a slight decrease in PESQ due to differences from the target speech, but it indicates that our model can also be effective on in-the-wild noisy speech.

\newpara{Inference speed.}
A significant highlight is the comparison of inference speeds as indicated in Table~\ref{table1}. We conduct experiments using the same RTX A4000 GPU device with an equally controlled situation. 
We compute the Real Time Factor (RTF) of each model which measures how much time is needed for the generation of one second of audio.
Our large size model exhibits an inference speed approximately 18 times faster than the diffusion-based model of the same size. This model's speed advantage is attributed to its ability to perform inference in just one step, whereas the diffusion-based model requires 30 steps to achieve sufficient performance. 
Moreover, we reduce the existing model size by more than half in our small size model. This lightweight model not only attains outstanding performances compared to other models but also accelerates inference speed about 22 times compared to AVDiffuSS.

\newpara{Sampling steps.}
As indicated in Fig.~\ref{fig_steps}, our model demonstrates robust performance across different sampling steps. 
This result denotes that unlike the diffusion-based approach \cite{lee2023seeing}, the 1-step inference is enough to get satisfying results in our model.
The flow matching objective is constructed based on an ODE, which learns a straighter path than a stochastic differential equation which is used in diffusion models. This attribute can enable fewer steps in inference compared to diffusion models.

\input{Tables/only_SS}
\input{Tables/prior}

\newpara{Speech separation.}
While speech enhancement focuses on improving the quality of the target speech in a noisy environment, speech separation aims to isolate the target speech from a mixture of multiple speeches. Given the shared objective of extracting the target speech, we evaluate our model by comparing it with other speech separation models, as summarized in Table~\ref{table_SS}. 
Among the baseline models, DiffSep~\cite{Diffsep} is an audio-only model that utilizes diffusion models, while VisualVoice~\cite{visualvoice} leverages multi-modal information without relying on generative models. 
Therefore, the concept of sampling steps, which is pertinent to generative models, does not apply to VisualVoice.

For this task, our large model is trained from scratch for 15 epochs to separate the target speech from a mixture of two speech signals from the VoxCeleb2 train set. 
Similar to the speech enhancement results, our model achieves comparable results in speech separation with 18 times faster inference than the previous state-of-the-art model~\cite{lee2023seeing}.

\newpara{Prior selection.}
In Table~\ref{table:prior}, we investigate the efficacy of the CFM approach based on a prior distribution $p_0$, serving as the initial point of inference. We select ${P}_{\theta}(\mathbf{y},\mathbf{f_v})$ as a mean of the prior distribution for CFM, which attains better performances in ESTOI and SI-SDR than setting the mean to zero.

%% file: Tables/only_SS.tex
\begin{table}
\centering
\caption{Speech separation results on the VoxCeleb2 test set. \textup{A-V} refers to the audio-visual model. The number of sampling steps does not apply to VisualVoice because it is not generative.}
\vspace{-1mm}
\resizebox{1.0\columnwidth}{!}{%
\begin{tabular}{lccccc}
\toprule
{Method}     & A-V  & Steps & PESQ $\uparrow$   & ESTOI $\uparrow$  & SI-SDR $\uparrow$   \\ 
\midrule
{DiffSep~\cite{Diffsep}}     &              & 30 & 2.202$\pm$0.004       &   0.595$\pm$0.013     &   ~3.971$\pm$0.436       \\
{VisualVoice~\cite{visualvoice}} & \checkmark     & N/A   & 1.953$\pm$0.001 & 0.765$\pm$0.001 & ~9.218$\pm$0.253   \\
{AVDiffuSS~\cite{lee2023seeing}}       & {\checkmark  }   & 30 & \textbf{2.520$\pm$0.007}   & \textbf{0.811$\pm$0.004} & \underline{11.852$\pm$0.418} \\

\midrule
{Ours}        & \checkmark   &1   & \underline{2.230$\pm$0.002} & \underline{0.796$\pm$0.002} & \textbf{12.263$\pm$0.006} \\

\bottomrule
\end{tabular}%
}
\label{table_SS}
\end{table}

%% file: Tables/prior.tex
\vspace{2mm}
\begin{table}
\centering
\caption{Ablation on the mean of prior distribution $\mu_0$ at time $t=0$ on VoxCeleb2 test set. Our model sets $P_{\theta}(\mathbf{y},\mathbf{f_v})$ as a mean of prior distribution. The result is compared with setting $\mu_0$ as $0$, which corresponds to the standard normal distribution.}
\vspace{-1mm}
\resizebox{0.87\linewidth}{!}{
\begin{tabular}{lcccc}
\toprule 
$\mu_0$  & PESQ $\uparrow$ &ESTOI $\uparrow$ &SI-SDR $\uparrow$  \\ 
\midrule
$0$            & \textbf{2.153$\pm$0.033} & 0.794$\pm$0.007 & 13.096$\pm$0.123\\
\midrule
$P_{\theta}(\mathbf{y},\mathbf{f_v})$    &  2.096$\pm$0.027    &    \textbf{0.796$\pm$0.006}      &    \textbf{13.370$\pm$0.111}        \\
\bottomrule

\end{tabular}
}
\label{table:prior}
\end{table}

%% file: Contents/Conclusion.tex
\section{Conclusion}
\label{sec:conclusion}
In this work, we propose FlowAVSE, an efficient audio-visual speech enhancement framework based on the conditional flow matching approach. Our model accelerates inference speed by approximately 22 times compared to the previous diffusion-based model. We design a light U-net architecture by removing repeated convolution modules, enabling fast inference while maintaining strong performance. The proposed model achieves state-of-the-art performance for audio-visual speech enhancement, particularly excelling in denoising metrics.

%% file: Contents/ack.tex
\section{Acknowledgements}
\label{ack}

This work was supported by the National Research Foundation of Korea (NRF) grant funded by the Korea government (MSIT, No. RS-2023-00222383).